\documentclass[12pt,preprint2]{aastex63}

\newcommand{\Ha}{H$\alpha$}

\newcommand{\Htwo}{H\textsubscript{2}}

\newcommand{\sbunits}{erg~s$^{-1}$~cm$^{-2}$~sr$^{-1}$}

\newcommand{\kms}{km~s{$^{-1}$}}

\newcommand{\ozone}{He$^{+}$+H$^{+}$}

\newcommand{\nzone}{He$\rm ^{o}$+H$^{+}$}

\newcommand{\tC}{{$\theta^1$~Ori~C}}

\newcommand{\hii}{\ion{H}{2}}
\newcommand{\hi}{\ion{H}{1}}

\newcommand{\oiii}{[\ion{O}{3}]}

\newcommand{\Cii}{[\ion{C}{2}]}

\newcommand{\nii}{[\ion{N}{2}]}

\newcommand{\Vscat}{{\bf V$\rm_{scat}$}}
\newcommand{\Vemit}{{\bf V$\rm_{emit}$}}
\newcommand{\Vscatnii}{{\bf V$\rm_{scat,[N II]}$}}
\newcommand{\Vscatoiii}{{\bf V$\rm_{scat,[O III]}$}}

\newcommand{\Vcomp}{{\bf V$\rm_{comp}$}}
\newcommand{\Scomp}{{\bf S$\rm_{comp}$}}

\newcommand{\Vlow}{{\bf V$\rm _{low}$}}

 \newcommand{\Vnewoiii}{{\bf V$\rm _{new,[O~III]}$}}

\newcommand{\Vnew}{{\bf V$\rm _{new}$}}
\newcommand{\Vpdr}{{\bf V$\rm _{PDR}$}}

\newcommand{\Vmifnii}{{\bf V$\rm_{mif,[N~II]}$}}
\newcommand{\Vmifoiii}{{\bf V$\rm_{mif,[O~III]}$}}
\newcommand{\Vnil}{{\bf V$\rm_{NIL}$}}

\newcommand{\Vmif}{{\bf V$\rm_{mif}$}}

\newcommand{\Vr}{{\bf V$\rm_{r}$}}

\newcommand{\Vlongoiii}{{\bf V$\rm _{long,[O~III]}$}}
\newcommand{\Vshortoiii}{{\bf V$\rm _{short,[O~III]}$}}
\newcommand{\Vlong}{{\bf V$\rm _{long}$}}
\newcommand{\Vshort}{{\bf V$\rm _{short}$}}
\newcommand{\Vlongnii}{{\bf V$\rm _{long,[N~II]}$}}
\newcommand{\Vshortnii}{{\bf V$\rm _{short,[N~II]}$}}
\newcommand{\Vred}{{\bf V$\rm _{red,[O~III]}$}}

\newcommand{\Vevap}{{\bf V$\rm_{evap}$}}
\newcommand{\Vevapnii}{{\bf V$\rm_{evap,[N~II]}$}}
\newcommand{\Vevapoiii}{{\bf V$\rm_{evap,[O~III]}$}}

\newcommand{\Sbothoiii}{{\bf S$\rm _{both,[O~III]}$}}
\newcommand{\Slongoiii}{{\bf S$\rm _{long,[O~III]}$}}
\newcommand{\Sshortoiii}{{\bf S$\rm _{short,[O~III]}$}}

\newcommand{\Slongnii}{{\bf S$\rm _{long,[N~II]}$}}
\newcommand{\Sshortnii}{{\bf S$\rm _{short,[N~II]}$}}

\newcommand{\Slowoiii}{{\bf S$\rm _{low,[O~III]}$}}
\newcommand{\Sscat}{{\bf S$\rm _{scat}$}}
\newcommand{\Sscatnii}{{\bf S$\rm _{scat,[N~II]}$}}
\newcommand{\Sscatoiii}{{\bf S$\rm _{scat,[O~III]}$}}

\newcommand{\cloud}{{\bf Orion-S Cloud}}
\newcommand{\ledge}{{\bf Ledge}}
\newcommand{\extledge}{{\bf Extended Ledge}}
\newcommand{\crossing}{{\bf Crossing}}

\newcommand{\outside}{{\bf Outside-Group}}

\newcommand{\lig}{{\bf Low-Ionization-Group}}
\newcommand{\nil}{{\bf NIL}}
\newcommand{\ave}{{\bf Ave}}

\newcommand{\NEsuper}{{\bf NE-Supergroup}}
\newcommand{\CRsuper}{{\bf Crossing-Supergroup}}
\newcommand{\SWsuper}{{\bf SW-Supergroup}}

\begin{document}

\title{Deciphering the 3-D Orion Nebula-III: Structure on the NE boundary of the Orion-S Imbedded Molecular Cloud}

\author{C. R. O'Dell\affil{1}}
\affil{Department of Physics and Astronomy, Vanderbilt University, Nashville, TN 37235-1807}

\author{N. P. Abel\affil{2}}
\affil{MCGP Department, University of Cincinnati, Clermont College, Batavia, OH, 45103}


\author{G. J. Ferland\affil{3}}
\affil{Department of Physics and Astronomy, University of Kentucky, Lexington, KY 40506}

\begin{abstract}

We have extended the work of Papers I and II of this series to determine at higher spatial resolution the properties of the embedded Orion-S Molecular Cloud that lies within the ionized cavity of the Orion Nebula and of the thin ionized layer that lies between the Cloud and the observer. This was done using existing and new \nii~(658.3 nm) and \oiii~(500.7 nm) spectra that map the central region of the Orion Nebula (the Huygens Region). However, it remains unclear how the surface brightness of the ionized layer on the Orion-S Molecular Cloud and that of a foreground Nearer Ionized Layer are linked, as the observations show they must be. It is shown that the Cloud modifies the outer parts of the Huygens Region in the direction of the extended hot X-ray gas.
\end{abstract}

 \section{Introduction}
 \label{sec:Intro}
 
This is the third of a series of papers [\citep{ode20a} (henceforth Paper-I) and \citep{ode20b} (henceforth Paper-II)] addressing structures within the Huygens Region of the Orion Nebula revealed by high-velocity-resolution spectroscopy. Using spatially large samples of spectra we determined in Paper-II that the region containing the Orion-S Cloud was characterized by widely changing strengths of velocity components in \oiii\ (500.7nm), a phenomenon most striking near a sub-region call the \crossing. In this paper we investigate this sub-region using higher spatial resolution groupings of spectra.

The \crossing\ and the Orion-S Cloud are especially important features as they lie in the direction of the X-ray bright \citep{gud08} portions of the nebula that are enclosed by the recently discovered \Cii\ Outer Shell \citep{pabst1,pabst2}. The Orion-S Cloud must interrupt the flow of stellar wind in that direction and also cast a radiation shadow. In Paper-IV \citep{ode20c} we will explore the Herbig-Haro flow designated as HH~269 that arises from within the Orion-S Cloud.

This paper differs from Papers-I and II in dealing with sequences of higher spatial resolution slit spectra of about 3\farcs6 to 4\farcs0 width and 8\arcsec\ to 13\arcsec\ length rather than averages of spectra over 10$\times$10\arcsec\ samples. Although the slit spectra have lower total signal-to-noise ratios, they are not blurred in spatial or spectral resolution by fine-scale structure in the nebula. We shall see that the sequences of spectra reveal patterns only hinted at in the large sample studies.

\subsection{Background of this study}
\label{sec:Background}

A line-of-sight ray out of the Orion Molecular Cloud (OMC) towards the observer first passes through a Photon Dominated Region (PDR) \citep{stacy,goi15}  and then the overlying Main Ionization Layer (MIL) predominately photo-ionized by \tC. The MIL is stratified into two zones of ionization, ordered by increasing distance from the \hii\ actual ionization front and the distance to the photoionizing star \tC. The zone closest to the ionization front  is composed of \nzone\ and emits the collisionally excited \nii\ (658.3 nm) line used in this study. The further zone is composed of \ozone\ and emits the collisionally excited \oiii\ (500.7 nm) line used in this study. Material from the PDR is continuously lost as the MIL gas expands as photoevaporation flow into the lower density regions further out. The expanding gas is accelerated away from the ionization front, with the result that the \oiii\ photoevaporation flow velocity should be greater than that of \nii. This photoevaporation flow has been well modeled by \citet{hen05} and are not affected by the global motions seen in layers of gas beyond \tC.

The region immediately around \tC\ is of lower density due to a stellar-wind blown bubble whose outer boundary is shown in projection 
by a High Ionization Arc \citep{ode20a}. Within the bubble is the Orion-S Molecular Cloud that is seen in absorption against the MIL radio continuum \citep{johnston83,mangum93,vdw13}. The observer's side of the Orion-S Molecular Cloud is illuminated by \tC , producing an ionized layer that is the optically brightest portion of the Huygens Region (Paper-II).  Proceeding outward it encounters a layer of ionized gas designated as the Nearer Ionized Layer (NIL) \citep{abel19}, Paper-I, and Paper-II. Further out still there are two layers of atomic gas and \Htwo\ \citep{abel16}. These were discovered in \hi\ 21-cm absorption \citep{vdw89} and more recently it
was found that one of these layers is part of an expanding outer shell covering the entire Extended Orion Nebula \citep{gud08} seen in \Cii\ 158 \micron\ emission \citep{pabst1,pabst2}.

In Paper-I and Paper-II we drew on the high-spectral-resolution Spectroscopic Atlas of Orion Spectra \citep{gar08} (henceforth `the Atlas'), compiled from a series of north-south spectra at intervals of 2\arcsec. The Atlas has a velocity resolution of 10 \kms\ and a seeing limited spatial resolution of about 2\arcsec. 
 We analyzed high signal-to-noise (S/N) spectra of \nii\ and \oiii , averaged over spatial boxes of 10\arcsec $\times$10\arcsec\ and groupings of these Samples into larger areas called Groups and Regions.  In Paper-I we established the large-scale properties of the Huygens Region (the well-studied bright region usually 
 identified with M42, the Orion Nebula), establishing that this region has a series of large-scale structures, the inner-most being the optically bright Main Ionization Front (MIF) and the outer-most being a Veil of atomic and molecular gas.
 
 In Paper-II we explored at low spatial resolution the region to the SW of the dominant ionizing star \tC , establishing that the area including the \cloud\ defies explanation by simple models.  The \lig\  was marked by very different  behaviors of radial velocities. In \oiii\ these velocities group at two values, contrary to the expectations of photoevaporation from an ionization front and is in contrast with the simpler behavior in \nii. 
  
In the current paper we target the \lig, which overlies the third star formation region in the Huygens Region, at higher spatial resolution. It lies 50\arcsec\ at 234\arcdeg\ from \tC\ and must be associated with the \cloud\ (it lies at the NE corner of the 21-cm \hi\ contour of the \cloud.
The \cloud\ is seen in radio absorption lines (hence it must lie in front of a source of radio continuum). The young stars lie on the east side of the Cloud and are the source of many collimated molecular and ionized outflows (jets).  Shocks associated with these jets span the Huygens Region. Extrapolation of the jets backwards 
gives their origins with varying degrees of accuracy. Good presentations of the radio sources and jets are in Fig.~1 of \citet{ode09} and Figures 15, 16, and 17 of 
\citet{ode08}. A more recent study \citep{ode15} refines the idea that although the sources are imbedded and not seen in the optical, they lie close enough to the PDR that many of their jets break out and become optically visible features.
This means that the geometry of the nebula is very different near this star formation region. 

\subsection{Nomenclature and adopted values}
\label{sec:nomenclature}

The list of terms and adopted values are presented in Paper-II, but additional terms are necessary in this study and an amended list is given below.

$\bullet$ Slit spectra are narrow rectangular areas of samples analogous to a short slit spectrum.

$\bullet$Profiles are the data from a series of slit spectra ordered along a single direction. 

$\bullet$ Adjacent Samples are a set of spectra selected to avoid the High Ionization Arc and the region surrounding the intersection of the Profiles called the \crossing.

$\bullet$ Samples are areas of 10\arcsec $\times$10\arcsec\ within which spectra from a spatially resolved atlas of spectra of certain emission-lines have been averaged. 

$\bullet$ Regions are groupings of Samples. 

$\bullet$ Sectors are groupings of Adjacent Samples,  Samples, and Regions grouped by orientation relative to the \crossing.

$\bullet$ The adopted distance is 388$\pm5$ \citep{mk17}. 

$\bullet$ The adopted velocity for the background PDR is \Vpdr\ = 27.3$\pm$0.3 \kms\ \citep{goi15}. 

$\bullet$ All velocities are expressed in \kms\ in the  Heliocentric system (Local Standard of Rest velocities are 18.1 \kms\ less). 

$\bullet$ Directions such as Northeast and Southwest are often expressed in short form as NE and SW.

\subsection{Outline of this paper}
\label{sec:outline}

In Section~\ref{sec:observations} we describe the division of the components of the resolved line profiles (Section~\ref{sec:VelSys}), how the rectilinear arrays of spatially resolved spectra from the Atlas were used to create series of spectra of a few arcseconds width at orientations useful for analyzing the Orion-S Cloud and groupings of the slit spectra into 'Adjacent Samples'  of higher signal to noise ratio (Section~\ref{sec:ObsSlits}. The location of the Adjacent Samples and their grouping  are described in Section~\ref{sec:OldSpec}. Profiles previously presented \citep{ode18} are presented in Section~\ref{sec:AnalProfiles}, now annotated for the current study. The methods of analysis of the spectra are descibed in Section~\ref{sec:methods}.An analysis of the profiles is presented in Section~\ref{sec:AnalProfiles}. A compilation of the data from Papers I and II, and this paper is given in Section~\ref{sec:compilation}. Section~\ref{sec:discussion} discusses the observational properties and their relation to the 3-D properties of the Orion-S Cloud and a summary of our conclusions and recommendations appear in Section~\ref{sec:conclusions}. Appendix~\ref{sec:DkArcProfile} presents a 
detailed study of an important small feature on the \cloud.

\section{Observations} \label{sec:obs}
\label{sec:observations}

\subsection{Characteristic Velocity Systems}
\label{sec:VelSys}

In Paper-I and Paper-II we have described how the spectra were created and de-convolved into velocity components. We repeat their description presented in descending velocity. \Vscat\ (ascribed to backscattering from dust particles in the background Photon Dominated Region (henceforth PDR)), \Vnew\  and \Vred\ (the former ascribed to material in the High Ionization Arc feature and the latter to material accelerated towards the host Orion Molecular Cloud by the stellar wind of \tC) or material that would normally be assigned to \Vlongoiii\ except for when the \Vshortoiii\ component is much stronger, \Vlongoiii\ (ascribed to emission from the ionized layer, the MIF, on the Orion Molecular Cloud (OMC) or the ionized layer of the \cloud\ facing the observer, and \Vshortoiii\ (usually weak and ascribed to a foreground Nearer Ionized Layer (the \nil) lying in the foreground of \tC).  These components 
are seen in both the \nii\ and \oiii\ emission-lines. In Paper-I and earlier studies the \Vshortnii\ and \Vshortoiii\ components were called \Vlow\ and the \Vlong\ components were called \Vmif. As in Paper-II, when discussed as emission from specific physical layers, the terms \Vmif\ and \Vnil\ are used. 

\subsection{Twin Strong Components are seen in Spectra within the \crossing}
\label{sec:twinLines}

In Papers-I and II, usually the strong MIF velocity component dominated each spectrum although in Paper-II we found Samples where \Vshort\ was stronger and \Vlong\ was not detected. Typical spectra and the short-comings of deconvolution of the weaker secondary components were discussed and illustrated \citep{ode18,ode20a}. Such spectra were encountered in the present study, but in the regions within the \crossing\ we often found that \Vshortoiii\ and \Vlongoiii\ components are both strong.

\begin{figure}
 \includegraphics
[width=\columnwidth]
 {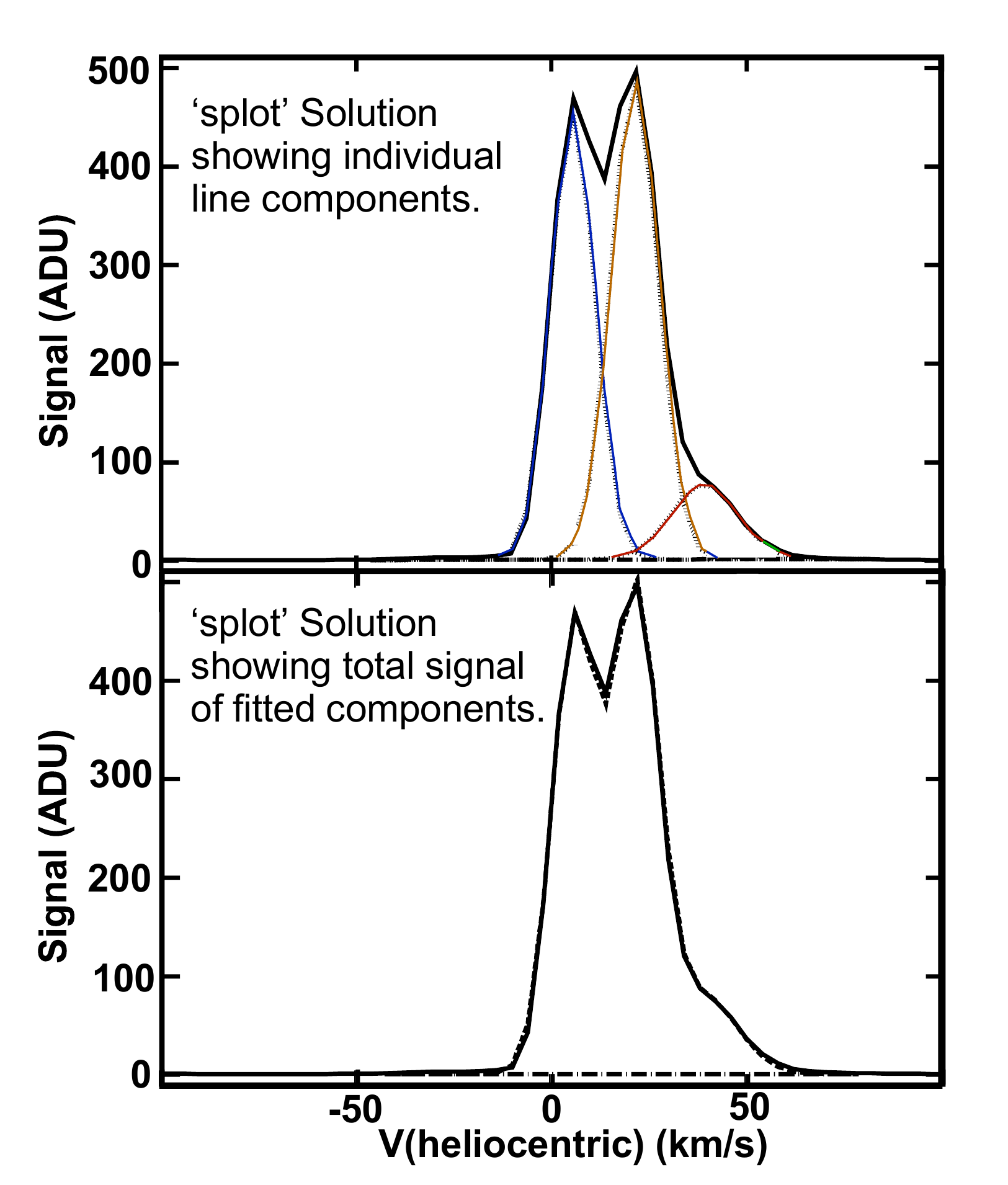}
\caption{This spectrum illustrates \oiii\ spectra in the middle of the \crossing\ region, as described in Section~\ref{sec:twinLines}. The top panel shows the observed line profile in black and the fitted components in color \Vshort -Blue, \Vlong -Orange, \Vscat -Red). In the lower panel, the observed line profile is in solid black and the composite of the three fitted components is a barely distinguishable dotted line.
The solutions using IRAF are: \Vshortoiii\ = 5.3 \kms\ (\Sshortoiii /\Slongoiii\ = 0.86), \Vlongoiii\ = 21.0 \kms (\Slongoiii\ = 1.00), \Vscatoiii\ = 39.1 \kms\ (\Sscatoiii /\Slongoiii\ = 0.23).
}
\label{fig:twinLines}
\end{figure}

In Figure~\ref{fig:twinLines} we show a good example of a twin strong component \oiii\ spectrum. This is spectrum 29 in the South-North Profile (Figure~\ref{fig:Profiles}) and was deconvolved using IRAF\footnote{IRAF is distributed by the National Optical Astronomy Observatories, which is operated by the Association of Universities for Research in Astronomy, Inc.\ under cooperative agreement with the National Science foundation.} task `splot'.

\subsection{Higher spatial resolution slit spectra}
\label{sec:ObsSlits}

We use small spatial samples as we study the \crossing, whereas in Papers I and II we used  10$\times$10\arcsec\  samples. This allowed an appropriately better spatial resolution in the \crossing . This was done using the same sequences of spectra as in \citet{ode18} and newly created N-S sequences of spectra.  

\subsubsection{Previously Used Slit Spectra Profiles}
\label{sec:OldSpec}
\begin{figure}
 \includegraphics
[width=\columnwidth]
 {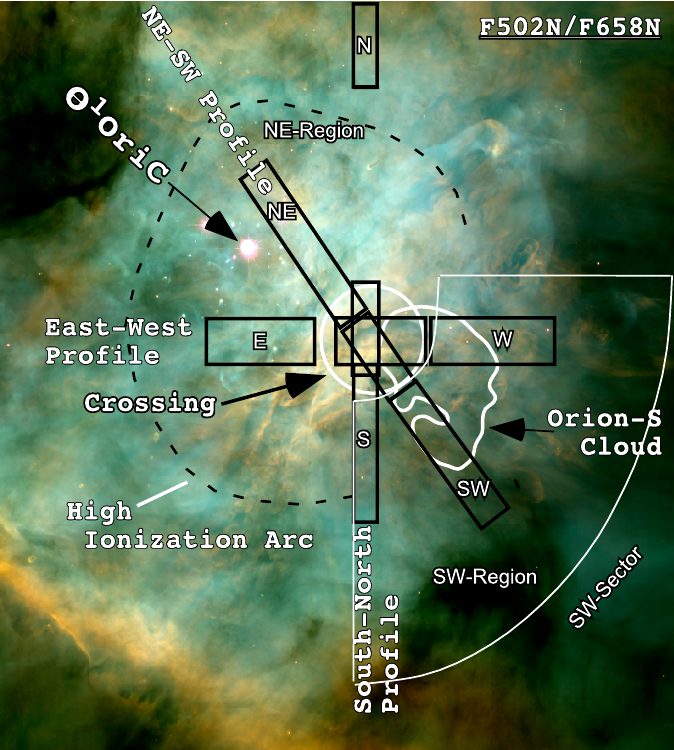}
\caption{This 194\arcsec$\times$216\arcsec\ (0.36$\times$0.40 pc) field of view (FOV) with north up and east to the left encloses the optically brightest part of the Huygens Region. It is centered 45\arcsec\ at PA (Position Angle) =215\arcdeg\ from \tC. The background image is from HST WFPC2 images coded by color (F658N, \nii \ Red; F656N, \Ha, Green; F502N, \oiii , Blue)  \citep{ode96}. It shows the location of the Adjacent and \crossing\ Samples discussed in Section~\ref{sec:ObsSlits}. The three \crossing\ samples for each profile are not named because of crowding but are shown. The \crossing\ circle has a diameter of 30\arcsec\ and is centered at 
5:35:13.95 -5:23:49.2 (2000). The SW-Sector white lines show the boundaries for those samples to the SW that exclude \crossing\ Samples. 
}
\label{fig:WideFOV}
\end{figure}

The first type of samples are taken from \citet{ode18} where a 
series of pseudo-slit spectra of approximately 3\farcs8$\times$8\arcsec --13\arcsec\  were gathered into congruent sequences called Profiles. Their locations are shown in Figure~\ref{fig:WideFOV} and the Profiles themselves in Figure~\ref{fig:Profiles}. These are the same as in \citet{ode18} except that they are now annotated for this study. Our analysis of these Profiles is quite different than in \citet{ode18} in that we now recognize the importance of the center of the \crossing\ and know when to group spectra into higher S/N ratio data.

 Some groups are within the \crossing\ (called the \crossing-Samples) and others (called the Adjacent-Samples) were selected to be close to the \crossing\ but avoided highly structured regions outside of the \lig . The Crossing-Samples do not overlap completely, thus their results need not be exactly the same, but collectively they are representative of conditions within the \crossing.

\begin{deluxetable*}{lccc}
\floattable
\setlength{\tabcolsep}{0.02in}
\tabletypesize{\scriptsize}
\tablecaption{Data from the Profile Spectra used in this study*
\label{tab:AllData}}
\tablewidth{0pt}
\tablehead{
\colhead{~~~~~~~~~~~~~}&
\colhead{~~~NE-Sector Adjacents} &
\colhead{~~~~~\crossing~~~~~} &
\colhead{~~~~~SW-Sector Adjacents}}
\startdata
\Vlongnii                    &19$\pm$1                 & 22$\pm$1                &19$\pm$1   \\
\Vshortnii                 &2.6$\pm$2.0             &4.6$\pm$2.0           & 0.8$\pm$2.4   \\
\Vscatnii                   & 33$\pm$2               & 38$\pm$2              & 32$\pm$1 \\
\Vscatnii\ - \Vlongnii   & 16$\pm$3               & 16$\pm$1                  & 16$\pm$1  \\
\Vlongoiii                       & 16.2$\pm$1.0        &19$\pm$2              &17$\pm$3\\
\Vshortoiii                   & 4.9$\pm$1.8         &   7$\pm$2              &6$\pm$1\\
\Vnewoiii                    & 27$\pm$2               & ---                                    & --- \\
\Vred                          &---                              & 19$\pm$2                   & 20$\pm$3\\
\Vscatoiii                    & 37$\pm$3                & 39$\pm$2             & 31$\pm$2 \\
\Vscatoiii\ - \Vlongoiii      & 20.9$\pm$3.7         & 20$\pm$3            & 18$\pm$3\\
\Slongnii /\Slongoiii $\dagger$          & 2.1$\pm$0.9      & 5.7$\pm$1.6        &---  \\
\Slongnii /\Sshortoiii           &26.5$\pm$18.1      &  5.4$\pm$1.2        & 3.4$\pm$1.8 \\
\Slongnii /)                      &  ---                        &  ---                                        & --- \\
~~~~(\Slongoiii + \Sshortoiii) &  1.94$\pm$0.67   &  2.9$\pm$0.6         &  1.7$\pm$0.3 \\
\enddata

~*~All velocities are Heliocentric velocities in \kms. 

**Samples group around these values.

$\dagger$\Slongnii /\Slongoiii ~ = 1.00 corresponds to a calibrated surface brightness (\sbunits) ratio  of 0.13.

$\dagger$$\dagger$ Without the highly scattered N-S Profile (19.6$\pm$9.2)  south adjacent values.

\end{deluxetable*}

Upon examination of the results for the Profiles, we identified three regions of similar data. The NE-Sector includes the North, NE, and East Adjacent-Samples, the \crossing\ contains the central Adjacent-Samples, and the SW-Sector contains the South, SW, and West Adjacent-Samples. Their locations are shown in Figure~\ref{fig:WideFOV}.

We present the averaged data of the multiple samples in Table~\ref{tab:AllData}.
Reading left to right one sees a progression of samples essentially flowing from the NE to the SW of the \crossing.
Vertically the results are ordered with the \nii\ data in the first five lines, the \oiii\ data in the next five, and the joint \nii\ and \oiii\ data on the last line.

\subsubsection{Newly Created Slit Spectra Sequences}
\label{sec:NewSpectra}
\begin{figure}
 \includegraphics
[width=\columnwidth]
 {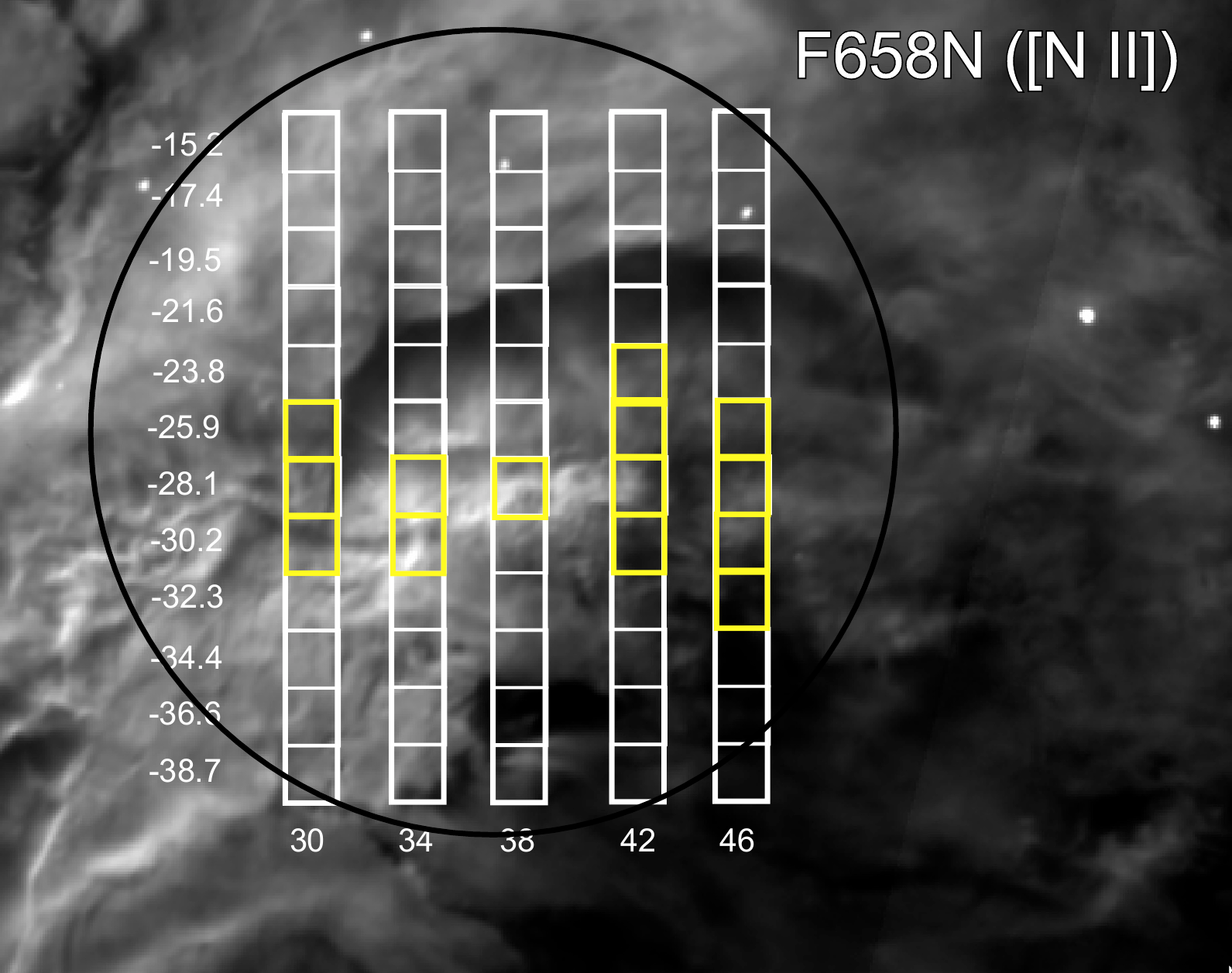}
\caption{
This 46\arcsec$\times$32\arcsec\ FOV is centered on the symmetry axis of the Dark Arc (within the \crossing\ which is shown with a dark circle). The boxes show the locations of the sampled spectra as discussed in Section~\ref{sec:discussion}. Those outlined in yellow define the \extledge\ samples, those to the north (up) the Dark Arc-North-Samples, and those to the south the Dark Arc-South-Samples. The columns are marked with the west displacement in arcsec from \tC\ and the individual north-south spectra in arcseconds south of \tC .The image was made with the HST WFPC2 camera in the \nii\ filter (F658N).}
\label{fig:DkArc658}
\end{figure}

\begin{figure}
 \includegraphics
[width=\columnwidth]
 {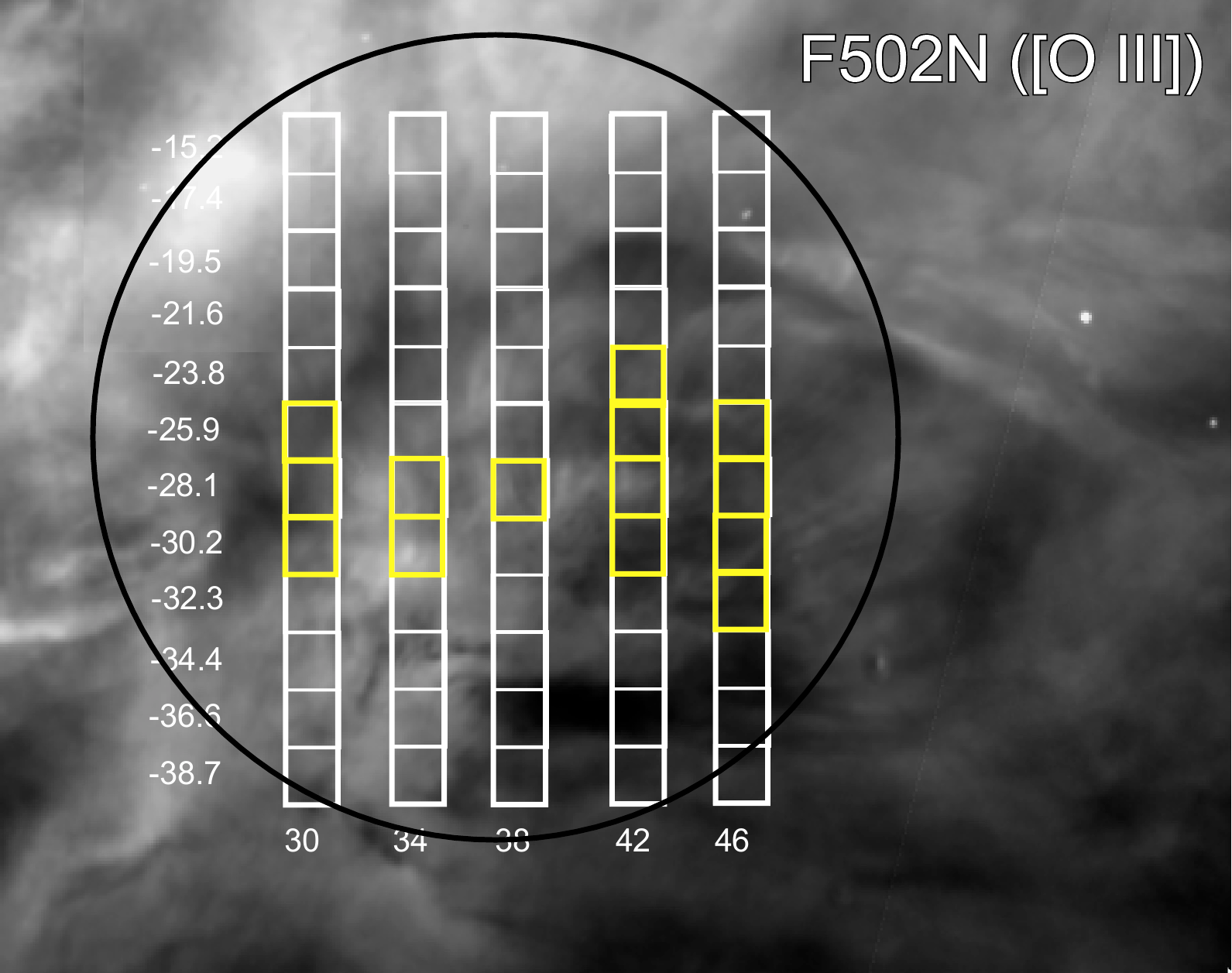}
\caption{
Like Figure~\ref{fig:DkArc658} except now showing the \oiii\ F502N image.
}
\label{fig:DkArc502}
\end{figure}

The second type of samples were created for this study. Within the \crossing , sequences of 12  2\farcs0 spectra at intervals of 4\arcsec\ in Right Ascension (RA) were made, as shown in Figures~\ref{fig:DkArc658} and \ref{fig:DkArc502}. This rectilinear array has the advantage of higher spatial resolution 
(2$\times$2\arcsec) and it better samples a nearly linear E-W feature in the middle of the \crossing . This is a well defined 'squiggly' feature, called here the \extledge , that was discussed in \citet{ode15} (where it was called the West-Jet). We determined membership in the \extledge\ using higher surface brightnesses and velocities as a guide and then derived averages of regions to the north and south of this (the Crossing-North and Crossing-South Spectra groups, henceforth North-Array and South-Array), with the results shown in Tables~\ref{tab:AllniiSources}, \ref{tab:AlloiiiSourcesVelocities}, and \ref{tab:AlloiiiSourcesSignals}. We believe that the disadvantage of averaging over varying features is appropriate because of the many samples in these groups.

Within the \extledge\ there is a \nii\ bright E-W linear
feature. Because of its lack of apparent motion in the plane of the sky this is not part of a moving jet. It is most likely to be a small portion of the MIF that is almost along the observer's line-of-sight. We now refer to this as the \ledge\ and it must be a small local escarpment with the higher side to the south. There are other features that are strong in \oiii\ on both sides of the \ledge\ with the ones on the west have measurable tangential motions. The \extledge\ and its location is shown in Figures~\ref{fig:DkArc658} and \ref{fig:DkArc502}. The ledge is pointed out in Figure~\ref{fig:Crossing}. 

The \ledge\ is discussed in detail in Appendix~\ref{sec:DkArcProfile} .

In addition to velocity components that fit into the usual classifications, very blue and weak \oiii\ velocity components were seen in a few of the 34\arcsec -West, 42\arcsec -West, and 46\arcsec -West profiles.


The data from these three regions are given within Tables~\ref{tab:AllniiSources}, \ref{tab:AlloiiiSourcesVelocities}, and \ref{tab:AlloiiiSourcesSignals}. In addition, results from Papers I and II are given for regions outside these samples.

\section{Analytic Tools} 
\label{sec:methods}

\subsection{Velocity variations when crossing tilted ionization layers}
\label{sec:deltaV}

A velocity profile across an escarpment such as the Bright Bar will produce a single velocity peak, followed by a decrease 
to the velocity in the region beyond the escarpment, as shown in Fig.~6 of \citet{ode18}.  When a profile crosses a discreet cloud, one would expect to see a double velocity peak, the first on the 
side illuminated by \tC, the other on the far side of the cloud. The magnitude of the velocity variations depend on \Vpdr\ and \Vevap\ (the photoevaporation flow velocity).  When the geometry is flat-on the observed \Vlong\ will be \Vpdr\ - \Vevap .
As the MIF becomes tilted the projection of the \Vevap\ components will be reduced and at an edge-on geometry one expects \Vlong\ = \Vpdr. 
If the tilt is not completely edge-on, the \Vlong\ variation will not be as great. 
As noted above, we would see this up and down velocity variation once in the profile of an escarpment and twice when the profile cross both sides of an isolated cloud.
The velocity pattern would be clearest when the profile cuts across the steepest part of the tilted front and would be less visible if the profile crosses it at
an angle. Likewise, because of the difference in the thickness of their emitting layers, we would expect to see the pattern better in the thinner \Vmifnii\ emitting layer than the thicker \Vmifoiii\ emitting layer. 

The Bright Bar is the best example of a single peaked velocity profile indicating that it is the edge of an escarpment and provides an aid when looking at similar features. The ionized portion shows signal peak and velocity maximum that is expected \citep{ode18} and infrared and radio studies of neutral gas shows the stratified presence of atoms and molecules consistent with and verifying PDR models \citet{tie93,goi16}.

The velocities and relative strengths of both the \nii\ and \oiii\ lines are presented in Table~\ref{tab:AllData}.

\subsection{Velocity differences of the \Vscat , \Vlong , and \Vshort\ components}
\label{sec:VelDif}

As noted previously (Paper-II) there is an expected relation between an emitting layer velocity (\Vemit) and its backscattered component velocity (\Vscat ) and \Vscat~-~\Vemit\  that is dependent upon the photoevaporation flow velocity (\Vevap), \Vpdr , and the tilt of the MIF. Under the assumption that the tilt is the same for the MIF and the \Vemit\ producing layer and that \Vevap\ is constant throughout the area being examined, a flat-on view would have \Vscat~-~\Vemit\ $\simeq$ 2$\times$\Vevap\ \citep{hen98} If the tilt increases towards an edge-on configuration, the line-of-sight component of \Vevap\ will decrease, producing a decrease in the observed \Vscat\ - \Vemit\ and an increase in the observed \Vemit. Application of this information discussed in Section~\ref{sec:PredObs}.

\subsection{Relative strength of the red-shifted backscattered component}
\label{sec:Albedo}

The ratio of signals (S) of the redshifted shoulder component of an observed line compared with a lower velocity component informs the question of what emission-line component is being backscattered. The observed ratio \Sscat /\Scomp\ must be much less that unity, reflecting the fact that the effective albedo must be low, unless there is a strong backscattering phase function. A large ratio means that this \Vcomp\ is not the source. If the source and the PDR are widely separated, the ratio will be unusually small.

\subsection{Surface Brightness variations}
\label{sec:SBexpectations}
For a photoevaporating ionization front the Surface Brightness (SB) in an \hi\ recombination line varies as the incident ionizing radiation \citep{bal91,agn06}. For a face-on flat MIF, this means that there would be a monotonic decrease in the SB at increasing distances (in the plane-of-the-sky) from the sub-\tC\ position. A concave MIF would have a slower decrease in SB with increasing angular distance and a convex MIF  would have a more rapid decrease. 
The general concave structure of the inner Huygen's Region is well established. Features within the Huygen's Region, such as the Bright Bar, are explained as steep rises in the MIF. The high SB there is explained by both the ionizing flux increasing due to the tilt and the fact that one is looking at the emitting layer edge-on.

More recently \citep{ode18}, the same geometry has been applied to explain why the brightest part of the nebula occurs to the NE of the Orion-S Cloud. Although the increase in surface brightness is certainly due in part to the tilt, the proximity to \tC\ of the NE boundary to the Orion-S Cloud must also be very important. In Paper-I we established that the distance of \tC\ from the MIF ionization boundary was in the range 0.1 -- 0.2 pc. The central value of 0.15 pc corresponds to 81\arcsec\ if projected onto the plane-of-the-sky. The peak surface brightness SB in \nii\ occurs at 33\arcsec\ (0.061 pc) from \tC\ in the plane-of-the-sky.  In Section~\ref{sec:Crossing} we pointed out that the Cloud is no closer to the observer than the plane including \tC\ or 0.05 pc (27\arcsec) beyond that plane. The corresponding range of physical distances between \tC\ and the NW edge of the Cloud is 0.061 -- 0.079 pc, both are closer to \tC\ than the \tC\ to the MIF ionization boundary. Even without consideration of the enhancement due to looking along an emitting layer seen edge-on, this explains why the SB in the Huygens Region is highest there. 

Of course the expectations for the SB becomes more complex when dealing with \nii\ and \oiii\ emission that occur in different zones within the ionized 
hydrogen layer, but even for their emission the first order expectations remain the same after consideration that the \nii\ 
emitting layer is thinner and closer to the actual ionization front than the \oiii\ emitting layer. However, if a region is lower ionization, then \nii\ emission will be enhanced relative to \oiii\ emission.

\section{Analysis of the Profiles}
\label{sec:AnalProfiles}	

\subsection{Analysis of the Profiles in and near the \crossing}
\label{sec:Crossing}
 The northeast side of the \cloud\ shows the 
velocity and ionization changes characteristic of an ionization front viewed edge-on \citep{md11,ode18,ode20b}, illuminated by \tC . The transition is shown in Figure~\ref{fig:WideFOV} where the \oiii\ dominated region transitions to an \nii\ dominated region along a SE-NW line. 
To the SW of this line lies the \cloud\ ionized on the observer's side by \tC. Our \crossing\ Samples are taken in the region of the Dark Arc feature (c.f. Section~\ref{sec:SBobserved}) and overlap the NE 21-cm absorption boundary of the Orion-S Cloud \citep{vdw13}. The \cloud\ is also seen in absorption in H$_{2}$CO \citep{johnston83,mangum93}. The velocities of these features are the same as the OMC in this direction \citep{tatematsu98,peng,tom16}. Since they are seen in absorption against an ionized gas continuum, the common interpretation is that this is a cloud lying within the main cavity of the nebula with portions of the MIF lying behind it.
In Paper-I we established that the \cloud\ lies at the same distance from the observer as \tC\ or no more than 0.05 pc beyond it.  

\begin{figure*}
\includegraphics
[width=7.5in]
 {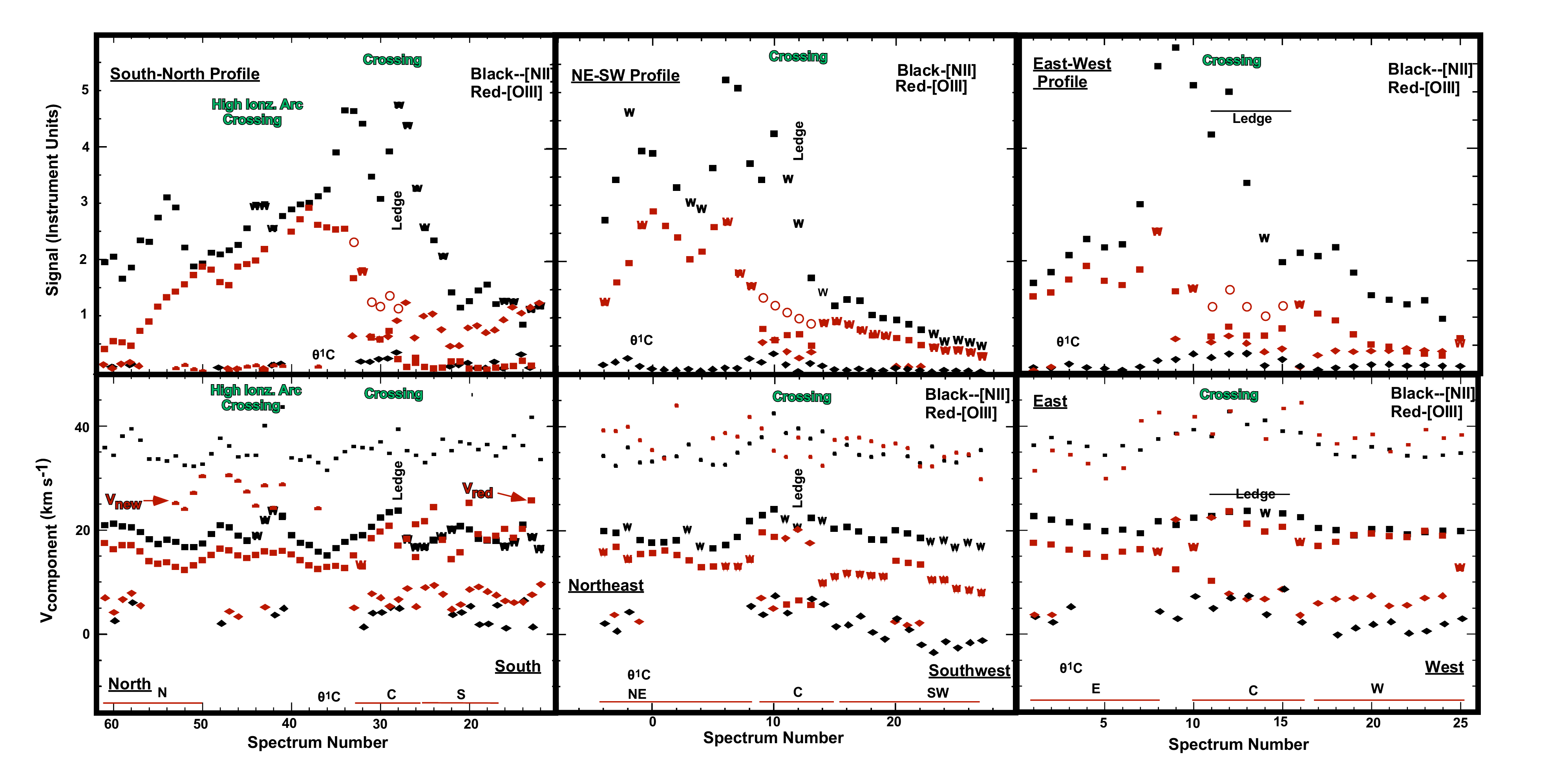}
\caption{Like Figures~10--12 from \citet{ode18} except annotated for this study. Large filled boxes long-components , small filled boxes scattered-components , horizontal boxes \Vnew\ and \Vred , and large filled diamonds short-components. Black indicates \nii\ and red \oiii. The spectra spacing is 3\farcs6 for the S-N and NE-SW profiles, and 4\farcs0 for the E-W Profile. \tC\ indicates the spectrum lying closest to \tC. Orion-S \crossing\ indicates spectra within the \crossing. The low red lines indicate the spectra included in the Adjacent Samples. The letter W indicates where the FWHM of \Vlongnii\ $\geq$ 18.0 \kms\ or when the FWHM of \Vlongoiii\ $\geq$ 16.0 \kms. The open red circles depict the sum of the \Sshortoiii\ and \Slongoiii\ components in the \crossing.
}
\label{fig:Profiles}
\end{figure*}

\subsection{Velocity Variations}
\label{sec:VelVar}

Given the guidelines described in Section~\ref{sec:deltaV} we can use the velocity variations in the profiles shown in Figure~\ref{fig:Profiles} to determine the geometry of the \crossing. 

\subsubsection{What do velocity variations in \nii\ tell us?}
 \label{sec:OriSniiVel}

Since the \nii\ emission comes from a thin layer close to an ionization front, it is not surprising the \Vlongnii\ changes demonstrate more continuous profiles than \Vlongoiii.  In this section we will discuss the \Vlongnii\ profiles and have always assigned the strongest component to \Vlongnii .

It has been established \citep{md11,ode18} that the Orion-S \crossing\ is immediately southwest of an ionization front viewed more nearly edge-on, like the Bright Bar. The question then becomes whether its velocity profiles show a single peak (hence it is an escarpment) or a double peak (hence it is a cloud illuminated on both sides. 

Each profile shows a slightly different variation in \Vlongnii, reflecting the fact that they trace different paths across the \crossing\ and nearby areas. The E-W Profile shows a broad peak in \Vlongnii\ occurring at spectrum 13 and a small local rise at spectrum 20, which lies on the east boundary of the {\bf Cloud's}  21-cm absorption boundary. Proceeding north in the S-N Profile shows two velocity peaks in \Vlongnii, the first at spectrum 28 (closest to a surface brightness maximum) and a second at spectrum 21 (outside of the \crossing, but well inside of the southern crossing of the High Ionization Arc so that it is not the source of this velocity peak). Further north there is a pair of velocity peaks (spectra 42 and 48) as one crosses the High Ionization Arc, indicating that this is a shell of material. The NE-SW Profile shows a peak in \Vlongnii\ near spectrum 10, with others at spectra 13 and 20. The peak at spectrum 13 may not be real as the surrounding spectra \Vlongnii\ lines are broad and may be blends of higher and lower velocity components. This would mean that there is the peak at spectrum 10, followed by a continuous reduction until the weak peak at spectrum 20.

Taken together, the \Vlongnii\ profiles indicate the crossing of a tilted ionization front near the center of the \crossing, followed by a flattening of the front, then crossing the outer boundary of this higher feature at about 30\arcsec\ from the center of the \crossing . 

It is surprising that crossing the edge of a raised feature (in the middle of the \crossing) occurs displaced from the transition region shown in Figure~\ref{fig:WideFOV}. It is as if this is a local structure superimposed on the broader \lig.

We can safely conclude that the profiles are across a raised feature (closer to the observer) because we see the expected local maximum in the surface-brightness (\Slongnii) at each of the first \Vlongnii\ peaks (crossing a depressed feature would produce \Slongnii\ minima because of shadowing of \tC\ radiation). 

The \Vshortnii\ behavior is harder to track because it is a usually a weak feature on the blue shoulder of the much stronger \Vlongnii\ component. The limitations of similar spectra are discussed quantitatively in \citet{ode18} and Paper-I. The lower signal to noise ratio of the smaller spectra as compared with the 10$\times$10\arcsec\ samples used in Papers I and II, means that we cannot now use the {\bf V$\rm _{long}$} data except where its signal becomes strong. These are discussed in Section~\ref{sec:ShortVel}. 

\subsubsection{What do velocity variations in \oiii\ tell us?}
 \label{sec:OriSoiiiVel}

Velocity variations in \oiii\ are much more complex that in \nii. The \Vshortoiii\ component is often strong and important, whereas in \nii\ it was always weak as compared with \Vlongnii. Examining the different \Vlongoiii\ profiles reveals the complex situation. The profiles in Figure~\ref{fig:Profiles} show components classified using the observed \Vr\ as a guideline, with the redder assigned to \Vlong\ and the bluer to \Vshort .

Proceeding from east to west in the E-W profile, we see that as one reaches the \crossing\ each spectrum suddenly has two significant components (\Sshortoiii$\simeq$\Slongoiii ) and they have two very different velocities. After passing through the \crossing\ \Sshortoiii\ becomes much less than \Slongoiii, but then increases and finally matches \Slongoiii , although the velocities of these two components remain about the same. \Vlongoiii\ has broad peaks at the same positions as \Vlongnii . The \Vshortoiii\ values are accurate (i.e. not questionable because of being a weak component on a stronger line's shoulder). This belief is reinforced by the fact that \Vshortoiii\ changes little as the components become of equal signal.

Proceeding south in the S-N Profile we again see two velocity peaks at spectra 42 and 48 as the shell of the High Ionization Arc is crossed. In this region we see higher velocity components (labeled \Vnew\ in Figure~\ref{fig:Profiles} that are probably associated with the Arc. Moving further south the velocity of \Vlongoiii\ changes dramatically upon entering the \crossing. Suddenly the stronger component \Vlongoiii\ drops to velocity values usually assigned to \Vshortoiii\ and remain there for the remainder of the profile. They are redder (20$\pm$3 \kms) and weaker components (labeled \Vred\ in Figure~\ref{fig:Profiles} that have a wide velocity dispersion and velocities similar to the \Vlongnii\ components (22$\pm$2 \kms ) in the other adjacent samples in the SW-Sector (Table~\ref{tab:AllData}). 
  
Proceeding SW in the NE-SW Profile we first see little \Vlongoiii\ structure until reaching the \crossing, at which point the \Vlongoiii\ and \Vshortoiii\ components become comparable signals. Results from spectra 12---18 are ambiguous because the strongest line is broad and the lower \Vlongoiii\ values there are in between the previous \Vlongoiii\ and \Vshortoiii\ values. The few narrow lines at spectra 20--22 fall at velocities usually associated with the \Vlongoiii\ and \Vshortoiii\ systems.

\subsubsection{What do the velocity differences tell us about the \crossing\ geometry?} \label{sec:OriSdifs}

Examination of \Vscat -\Vlong\ in Table~\ref{tab:AllData} indicates that for the backscattered \nii\ component, the well-defined difference of 16 \kms\ is close to the expected value of  14$\pm$2 \kms\ for a flat-on viewing angle.  

The difference for \oiii\ may change slightly, as the line-of-sight moves from the NE-Sector Adjacent-Samples (21$\pm$4 \kms ), through the \crossing\ (20$\pm$3 \kms ), to the SW-Sector Adjacent-Samples (18$\pm$3 \kms ). All of these are unimportantly lower than the predicted range 20$\pm$6 \kms.

The agreements of observed and expected velocity differences indicate that the MIF emission is in fact being backscattered by the nearby underlying PDR. The smaller dispersion of the \nii\ values indicates that there is not a big change in the viewing angle in the three regions. This is consistent with our intentionally not including the region of known high tilt on the NE boundary of the Orion-S Cloud. The same can be said for the \oiii\ emission, although with less certainty because of the larger probable errors.

\subsubsection{Observed Surface Brightness variations}
\label{sec:SBobserved}

The geometry of the nebula that produced the velocity variations described in Sections~\ref{sec:OriSniiVel} and \ref{sec:OriSoiiiVel} should also produce variations in the apparent surface brightness (SB) of the nebula. In this section we examine the SB variations using the signal in the Atlas as in \citet{ode18}, where the conversion to power units is also explained. 

None of the profiles cross through \tC , but their minimum distances (the \tC -tangent) from that star should be marked by a local peak in SB. These \tC -tangent points (spectrum 36 for the S-N Profile, spectrum -1 for the NE-SW Profile, spectrum 3 for the E-W Profile) are shown in Figure~\ref{fig:Profiles}.
In the absence of structure in the region covered by the profiles, we would expect to see a single peak in the SB at the \tC -tangent location with a monotonic drop in the SB in both directions away from this point. We see peaks in both ions near each of the \tC -tangent points and a continuous drop into the NE-Sector, interrupted only in the S-N profile where it crosses the high ionization arc and an even more removed feature to the north, seen only in \Slongoiii . 
Variations in \Vlongnii\ and \Vlongoiii\ accompany the High Ionization Arc passage. 

We also see variations in the SB in both ions as one traces outward (into the SW-Sector) from the \tC -tangent point.  If these profiles simply traced across an escarpment, one would expect to see a monotonic rise to a SB maximum as the observed {\bf V$\rm _{long}$} increases.  
This should be more obvious in \nii\ because its emitting layer is thinner.  The passage of the escarpment accounts for the \Slongnii\ peaks in each profile at 
S-N Profile (spectrum 34), NE-SW Profile (spectrum 7), and E-W Profile (spectrum 9). Small increases in \Vlongnii\ can be attributed to the passage across the 
escarpment.  

It should also be noted that unlike \nii , the sum of the \Slongoiii\ and \Slowoiii\ components in the \crossing\ approximate the total \Slongoiii\  value needed to provide a smooth interpolation of the SB over the \crossing\ as indicated by the open red circles in Figure~\ref{fig:Profiles}. This is consistent with the dark features being less conspicuous in \oiii. The comparable strength of both components makes their measurement more accurate than when \Slowoiii /\Slongoiii\ is low. The fact that the low components are strong raises the possibility that in the \crossing\ the \Vshortoiii\ emitting layer is competing with the \Vlongoiii\ emitting layer for the higher ionization energy photons necessary to produce \oiii\ as discussed in detail in Paper-II.        

\begin{figure}
 \includegraphics
[width=\columnwidth]
 {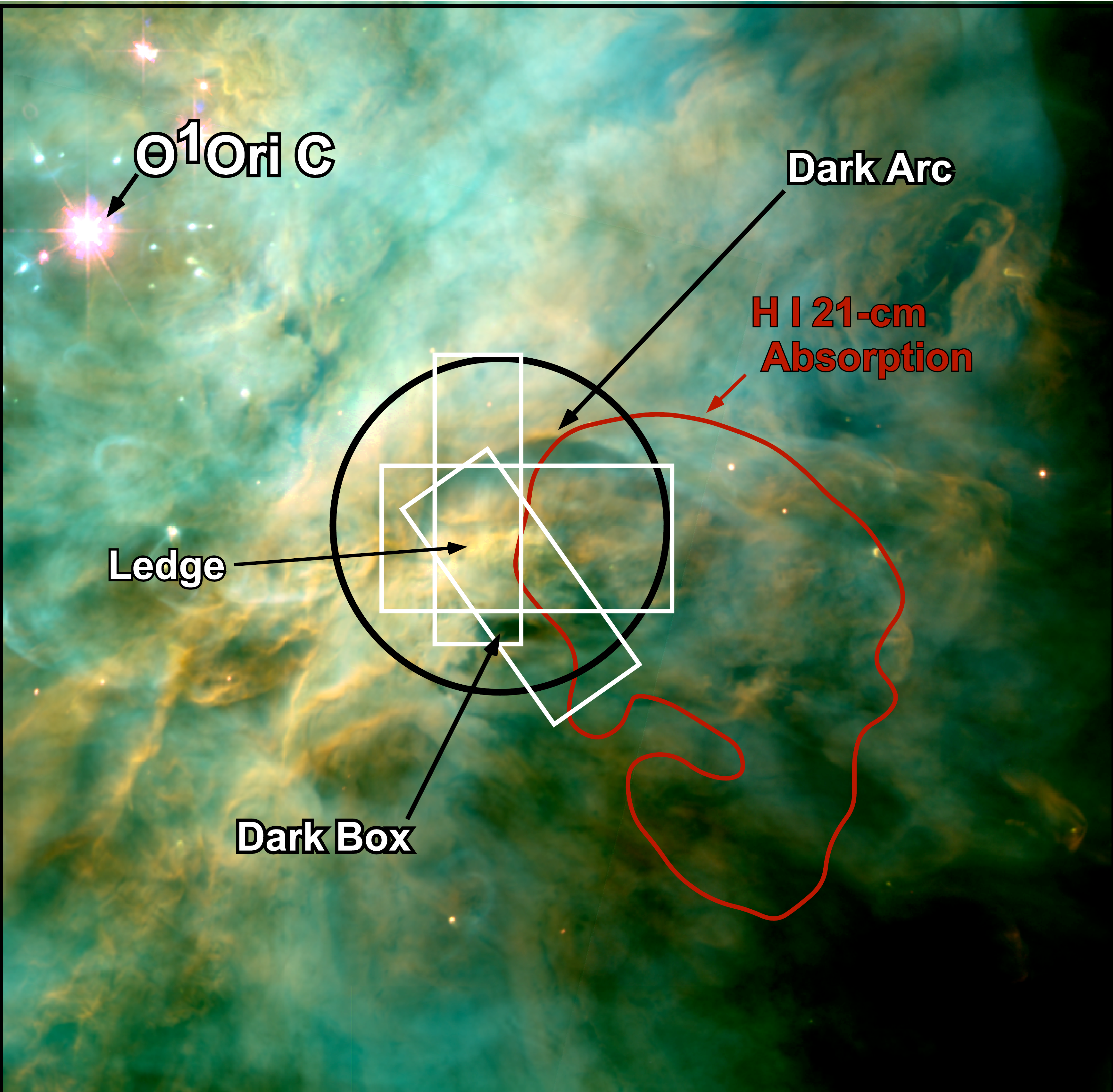}
\caption{
This 100\arcsec$\times$98\arcsec\ (0.19$\times$0.18 pc) FOV is centered on the \crossing\ (black circle) region and encloses the optically brightest part of the Huygens Region. The two small low surface brightness features (Dark Arc and Dark Box) discussed in Section~\ref{sec:SBobserved} are labeled. The background image is from HST WFPC2 images coded by color (F658N, \nii \ Red; F656N, \Ha, Green; F502N, \oiii , Blue)  \citep{ode96}.
The outer \hi\  absorption contours with heavy red lines are from \citep{vdw13}. The white rectangles indicate the location of the Central sections of the three profiles.}
\label{fig:Crossing}
\end{figure}

The feature called here the \extledge\ appears to be a small escarpment region, with the south side being higher. 
It is denoted in Figure~\ref{fig:Crossing}. Although it looks like a jet, there are 
no tangential velocities \citep{ode15} on its east end but several in \oiii\ and a few in \nii\ beginning at its west end. It is studied in depth in Paper-IV.

The point of passage across this east-west oriented feature (Figure~\ref{fig:Crossing} is marked by the word \ledge\ in each panel of Figure~\ref{fig:Profiles}. It is drawn with a line in the E-W Profile because the profile passes along the axis of the feature. At each marked position there is an associated local peak in \Slongnii , attributable to the feature.  Similarly, there is an increase in \Vlongnii\ at these locations. This important feature is discussed further in Appendix~\ref{sec:DkArcProfile}.



It is probable that a feature known as the Dark Arc \citep{ode00,ode15} also plays a role in the SB profiles. 
In Figure~\ref{fig:Crossing} we present our best resolution color image. The Dark Arc and Dark Box low SB features are seen in both ions, especially \nii . They are discussed in detail in Section~3.2.1 of \citet{ode15}. Although they are not understood, they are thought to be regions of low emissivity near the ionized layer on the observer's side of the Cloud, possibly small scale escarpments produced by one or more of the many stellar outflows in this region.
In the S-N and NE-SW Profiles the Dark Arc feature occurs at the SB minimum. In the E-W Profile the dip occurs at the spectrum
including the east boundary of the Dark Arc, where it is near aligned nearly N-S.


Examination of the \oiii\ profiles shows that the clear division into \Vshortoiii\ and \Vlongoiii\ components begins in the same samples that dip in \nii\ SB.  
The \Vlongoiii\  SB drops to lower than interpolation from adjacent spectra and the new \Vshortoiii\ spectra are of comparable strong to the \Vlongoiii\ components.

{\bf Most significant is that the total strength of the \Vlongoiii\ and \Vshortoiii\ components fall onto a smooth interpolation of adjacent values.}

\section{A compilation of data for the large scale features and groups}
\label{sec:compilation}
Averaged results for the samples described above and nearby regions from Papers I and II are given in Tables~\ref{tab:AllniiSources}, \ref{tab:AlloiiiSourcesVelocities}, and \ref{tab:AlloiiiSourcesSignals} In addition to the results of this study. The arrangement (top to bottom rows) is basically from NE to SW.

We have not used the results from the SW-Region discussed in Paper-I because it overlaps with multiple smaller groups identified here and in Paper-II. 
Two groups rendered in italics are affected by the slightly larger FWHM of the \Vlongoiii\ component . In those (the \lig\ and the \outside)
this width means that there is no chance of seeing a weak companion component and their values must be used cautiously.

Where there are two lines of entries for a group in \oiii , the property that showed the biggest division in entries
is marked ***. The composition of those subgroups was used to derive the other characteristics. When there was no
obvious difference in the two groups, a single value is given. The line containing the higher \Vlongoiii\ subgroup is always located higher in Tables~\ref{tab:AlloiiiSourcesVelocities} and \ref{tab:AlloiiiSourcesSignals}.

\section{Discussion}
\label{sec:discussion}

\begin{deluxetable*}{lcccc}
\floattable
\setlength{\tabcolsep}{0.02in}
\tabletypesize{\scriptsize}
\tablecaption{Data from All \nii\ Sources*
\label{tab:AllniiSources}}
\tablewidth{0pt}
\tablehead{
\colhead{Group Name}&
\colhead{\Vlongnii} &
\colhead{\Vshortnii} &
\colhead{\Vscatnii -\Vlongnii} &
\colhead{~~\Sscatnii /\Slongnii}}
\startdata
NE-Region           & 22$\pm$2    & 6$\pm$4  & 18$\pm$4  & 0.06$\pm$0.03 \\
Inside-Group        & 22$\pm$3    & 3$\pm$1  & 17$\pm$2  & 0.07$\pm$0.05 \\
NE-Sector            & 19$\pm$1    & 3$\pm$2  &  16$\pm$3 & 0.11$\pm$0.04 \\
{\bf Low Ioniz.-Group} &  21$\pm$2   & 6$\pm$3  &  16$\pm$1  & 0.08$\pm$0.03 \\
{\bf North-Array}          &  22$\pm$1   & 5$\pm$2  & 16$\pm$1 &  0.13$\pm$0.05 \\
\extledge              &  25$\pm$2   & 7$\pm$2   & 17$\pm$2 & 0.08$\pm$0.05 \\
{\bf South-Array}         &  24$\pm$1   & 5$\pm$2   & 19$\pm$2 & 0.04$\pm$0.02 \\
SW-Sector           &  19$\pm$1   & 1$\pm$2   & 16$\pm$1 &  0.13$\pm$0.03 \\
Outside-Group     &  19$\pm$2   & 3$\pm$3  & 16$\pm$1 &  0.13$\pm$0.04 \\
\enddata
~~~

~*~All velocities are Heliocentric velocities in \kms. 

\end{deluxetable*}
\begin{deluxetable*}{lccccc}
\floattable
\setlength{\tabcolsep}{0.02in}
\tabletypesize{\scriptsize}
\tablecaption{Data from All \oiii\ Sources-Velocities*
\label{tab:AlloiiiSourcesVelocities}}
\tablewidth{0pt}
\tablehead{
\colhead{Group Name}                            &\colhead{~~\Vlongoiii} &\colhead{\Vshortoiii} &\colhead{\Sshortoiii/\Slongoiii}  &\colhead{\Vscatoiii-\Vlongoiii} &\colhead{\Vscatoiii-\Vshortoiii} }
\startdata
NE-Region                                        & 18$\pm$2                    & 8$\pm$2                 & 0.10$\pm$0.03                         &19$\pm$4                               &    ---                       \\
Inside-Group                                     & 16$\pm$2                    & 3$\pm$1                 & 0.05$\pm$0.03                         &18$\pm$4                               &      ---                      \\
NE-Sector                                         & 16$\pm$2                    & 5$\pm$2                 & 0.04$\pm$0.09                         &21$\pm$4                               & ---                         \\
{\it Low Ioniz.-Group}$\dagger$$\dagger$&{\it13$\pm$2(6)}       & {\it8$\pm$2(8)}             &---**                                  &{\it21$\pm$3(6)}                          & {\it29$\pm$5(5)}        \\
{\bf North-Array}                                &  21$\pm$2(7)***          & 6$\pm$1(7)             & 0.8$\pm$0.2(7)                        &19$\pm$1(7)                           &35$\pm$2(15)            \\
{~~\bf North-Array}                            &  16$\pm$1(11)***        & 4$\pm$1(8)             & 0.2$\pm$0.1(8)                        &23$\pm$2(8)                           &---           \\
\extledge                                            &  20$\pm$2                  & 6$\pm$2                 & 1.0$\pm$0.3                             &19$\pm$2                               &34$\pm$2(14)         \\
 {\bf South-Array}                              &  18$\pm$2                   & 7$\pm$2                 & 3.2$\pm$1.0(7)***                    &16$\pm$1                                &28$\pm$3(20)          \\
~~{\bf South-Array}                           &                                     &                                & 0.9$\pm$0.2(9)***                    &21$\pm$1                              &---           \\
SW-Sector                                    & 20$\pm$3(9)               &8$\pm$2(9)               &10.3$\pm$1.9(8)***                  &15$\pm$2(9)                          &27$\pm$4(9)             \\
~~SW-Sector                                        &  17$\pm$2(11)             & 5$\pm$2(11)           &1.0$\pm$0.7(13)***                   &19$\pm$2(11)                         &31$\pm$2(11)             \\
{\it Outside-Group} $\dagger$$\dagger$&    ---                             &{\it11$\pm$2}          & {\it$>$$>$1}                                 &  ---                                          &{\it24$\pm$2}                  \\ 
\enddata
~~~

~*~All velocities are Heliocentric velocities in \kms.  

**\Vlongoiii\ and \Vshortoiii\ were not found in the same samples.

***Samples group around these values.

$\dagger$\Slongnii /\Slongoiii ~ = 1.00 corresponds to a calibrated surface brightness (\sbunits) ratio  of 0.13.

$\dagger$$\dagger$Results are affected by the unusually large FWHM of the \Vlongoiii\ components.

\end{deluxetable*}

\begin{deluxetable*}{lcccccc}
\floattable
\setlength{\tabcolsep}{0.02in}
\tabletypesize{\scriptsize}
\tablecaption{Data from All \oiii\ Sources-Signal Ratios*
\label{tab:AlloiiiSourcesSignals}}
\tablewidth{0pt}
\tablehead{
\colhead{Group Name}                        &\colhead{~~\Sscatoiii } &\colhead{\Sscatoiii } &\colhead{\Sscatoiii}  &\colhead{\Slongnii }  &\colhead{\Slongnii }   &\colhead{\Slongnii } \\
\colhead{}                                      &\colhead{/\Slongoiii}     &\colhead{/\Sshortoiii} &\colhead{/\Sbothoiii}&\colhead{/\Slongoiii}&\colhead{/\Sshortoiii}&\colhead{/\Sbothoiii$\dagger$}}
\startdata
NE-Region                                     &0.06$\pm$0.03            &                ---              &        ---                    &1.5$\pm$0.5            &           ---                    &--- \\
Inside-Group                                  &0.07$\pm$0.02            &        ---                      &      ---                     &1.3$\pm$0.1(15)      &    ---                           & --- \\
NE-Sector                                      &0.07$\pm$0.03(31)      &       ---                       &         ---                   &2.1$\pm$0.9            &         ---                      & --- \\
{\it Low Ioniz.-Group}$\dagger$$\dagger$&{\it0.12$\pm$0.04(6)}&{\it0.14$\pm$0.07(7)}&    ---**            &{\it2.8$\pm$0.3(6)}   &{\it3.1$\pm$0.4(6)}          &---** \\
{\bf North-Array}                             &0.16$\pm$0.07(7)        &0.19$\pm$0.05(7)     &0.10$\pm$0.04(10) &4.6$\pm$1.3(10)     &5.6$\pm$1.2(7)           &2.5$\pm$0.5(7) \\
{~~\bf North-Array}                         &0.14$\pm$0.04(8)        &0.7$\pm$0.3(8)        &0.10$\pm$0.03(10) &2.5$\pm$0.6(8)       &13.3$\pm$18.0(8)       &2.1$\pm$0.4(10) \\
\extledge                                         &0.26$\pm$0.12            &0.24$\pm$0.08         &0.13$\pm$0.04       &4.9$\pm$1.2(10)***      &5.0$\pm$1.0(11)         &3.0$\pm$0.9  \\
~~\extledge                                     &                                    &                                 &                               &15.4$\pm$5.4(3)***     &0.88$\pm$0.5(2)          &                      \\
{\bf South-Array}                             &0.6$\pm$0.3(7)            &0.14$\pm$0.05(6)  &0.12$\pm$0.03(6)      &$>$$>$1                        &2.7$\pm$0.5(6)       &2.5$\pm$0.6(6)  \\
~~{\bf South-Array}                         &1.1$\pm$0.3(10)          &0.20$\pm$0.08(10)   & 0.14$\pm$0.05(9)  &$>$$>$1                        &3.8$\pm$1.1(10)     &3.0$\pm$0.9(10) \\ 
SW-Sector                                  &1.2$\pm$0.6(11)          &0.16$\pm$0.10(9)     &0.13$\pm$0.06(9)  &19.5$\pm$8.8(9)       &2.3$\pm$0.4(9)                  &2.0$\pm$0.3(9)                       \\       
~~SW-Sector                                      & 0.20$\pm$0.06(9)       &0.33$\pm$0.11(11)   &0.12$\pm$0.03(11) &2.7$\pm$0.7(10)      &4.6$\pm$1.6(11)         &1.5$\pm$0.1(11)  \\
{\it Outside-Group}$\dagger$$\dagger$& ---                               &{\it0.13$\pm$0.04}         &---                           &  ---                             &{\it1.6$\pm$0.2}                       &---  \\ 
\enddata

~~~

~*~All velocities are Heliocentric velocities in \kms. 

**\Vlongoiii\ and \Vshortoiii\ were not found in the same samples.

***Samples group around these values.

$\dagger$\Slongnii /\Slongoiii ~ = 1.00 corresponds to a calibrated surface brightness (\sbunits) ratio  of 0.13.

$\dagger$$\dagger$Results are affected by the unusually large FWHM of the \Vlongoiii\ components.

\end{deluxetable*}

The three compiled tables of data can be used to investigate the structure and properties of a NE-SW swath across the Huygens Region. 

Examination of the results in Tables~\ref{tab:AllniiSources}, \ref{tab:AlloiiiSourcesVelocities}, and \ref{tab:AlloiiiSourcesSignals} illustrates how the unusual behavior of the \oiii\ velocity components found over a large area in Paper-II are first encountered in the \crossing , and that within the \crossing\ the behavior originates near a narrow east-west feature, the \extledge . To the NE the \Vshortoiii\ component is weak as compared with \Vlongoiii , but this is reversed as one passes the \crossing. 

These results refine the discovery in Paper-II that the \Vshortoiii\ component appeared in five of the nine large samples that included all or part of the \crossing . In addition, the \Vshortoiii\ component appeared without a \Vlongoiii\ component in fifteen other Samples to the south and southwest of the \crossing. This indicates that although the two velocity systems may originate in the \crossing, they occur throughout the \lig .

In contrast, the \Vshortnii\ component is always weak as compared with \Vlongnii. In Table~\ref{tab:NIL} we see that 
\Sshortnii /\Slongnii\ is 0.14, 0.25, and 0.11 in the three Supergroups. 

We see in these tables that the \oiii\ properties within samples often break down into groupings (subgroups), reflecting the fact that within the scale 
of the features there are a variety (at least two) of characteristics. The subgrouping begins in the North-Array and extends into the SW-Sector.

The data in Tables~\ref{tab:AllniiSources}, \ref{tab:AlloiiiSourcesVelocities}, and \ref{tab:AlloiiiSourcesSignals} can also be used to derive other properties of the regions sampled, as shown in the following sub-sections.

\subsection{Supergroups and the Relation of the short velocity features to the \nil}
\label{sec:ShortVel}

\begin{deluxetable*}{lccc}
\floattable
\setlength{\tabcolsep}{0.02in}
\tabletypesize{\scriptsize}
\tablecaption{Data for Supergroups*
\label{tab:NIL}}
\tablewidth{0pt}
\tablehead{
\colhead{Property}&
\colhead{\NEsuper }&
\colhead{\CRsuper } &
\colhead{\SWsuper } 
}
\startdata
\Vlongnii                     &  21$\pm$1                       & 24$\pm$1                    & 19$\pm$1 \\
\Vshortnii                    &  3$\pm$2        & 6$\pm$3                      & 2$\pm$2   \\
\Vlongoiii                     &   19$\pm$3                     & 21$\pm$2                    &  18$\pm$3  \\
\Vshortoiii                   & 4$\pm$2             &6$\pm$2                     & 6$\pm$2    \\    
\ave\Slongnii                      &171.3$\pm$90      &412.5$\pm$157         & 129$\pm$27 \\
\ave\Sshortnii                     & 13.8$\pm$4.4       &37.5$\pm$19         & 13.8$\pm$3.8 \\
\ave\Slongoiii                      & 108.3$\pm$47    &71.4$\pm$15           & 30.2$\pm$20.2 \\
\ave\Sshortoiii                     & 8.98$\pm$5.3       &62.1$\pm$9       &49.8$\pm$23.2 \\  
\ave\Slongnii /\ave\Slongoiii  &1.6      &5.8         & 4.3 \\
\ave\Slongnii /\ave\Sshortoiii  &19.1      &6.6            & 2.6 \\
\enddata
~*~All velocities are Heliocentric velocities in \kms.  

**Samples group around these values. Parentheses enclose the number of samples within each subgroup. 
\end{deluxetable*}

The Nearer Ionized Layer (the NIL) lies across much of the Huygens Region and is best characterized in Paper-I, where its velocity and relative strength is discussed and modeled, using data from near \tC.
It is appropriate to see if the \Vshort\ components we examine in this study are part of the NIL or have been altered by the conditions on the ionized layer atop the Orion-S Cloud. We have done this using 
the velocities and signals.

We gathered the groups into 'Super-Groups' of \NEsuper\ (NE-Region, Inside-Group, NE-Sector), \CRsuper\ (North-Array, \extledge , South-Array), and \SWsuper\  
(SW-Sector). Within each Supergroup we created a subgroup where velocities and signals existed for all components. 
Within these subgroups we determined the average \Vshortnii\ and \Vshortoiii, with the results shown in Table~\ref{tab:NIL}.

We also created Average signals for the subgroups in both the long and short components of both ions. These are preceded with \ave\ in Table~\ref{tab:NIL} and are more accurate measures of the surface brightness, whereas signals over samples are a mix of features from different parts of the samples. The $\pm$ symbols do not indicate uncertainties, rather, they are the 1-$\sigma$ spreads of the samples. The averaged values are used in our discussion of ionization changes (Section~\ref{sec:ionization}.

\Vshortnii\ shows the widest variation in value between the Supergroups. The \CRsuper 's value of 6$\pm$3 is slightly higher than the interpolation (2.5$\pm$2) between the adjacent Supergroups. If this marginal evidence is accepted, then it is an indication that the \nii\ emitting layer of the NIL that lies in from of the \cloud\ has been affected, but in the opposite sense expected from photoevaporation flow from the ionized surface of the \cloud\ (that would be a more negative \Vshortnii). In our discussion of the full profiles passing over the \crossing\ (Section~\ref{sec:OriSniiVel}), we saw that this region's \Vlongoiii\ values indicate that it is a raised region, which is consistent with its being part of the ionized layer on the observer's side of the \cloud. If the higher value of  \Vshortnii\  is accepted it could be due to interactions with the \cloud 's surface, thus arguing for a small separation of positions. In this case the more positive velocity would have to arise from flow away from the NIL and the observer. 

We see that the \Vshortoiii\ is essentially constant across the Supergroups at 5$\pm$2 \kms. This places it at a characteristic velocity of the NIL and argues that the velocity of the \oiii\ component of the NIL is not affected by the \cloud. 



\subsection{Changes of ionization in and near the \crossing}
\label{sec:ionization}



Since we see remarkable changes in the \Sshortoiii /\Slongoiii\ ratio as one moves from the NE to the SW it is important to determine if these are due to one component increasing while the other decreases, or vice versa. This is best done with the \ave\ signals given in Table~\ref{tab:NIL}. 
 
There we see that \ave\Slongnii\ has decreased more than a factor of 0.75 when going from the NE to the SW side of the crossing, while \ave\Slongoiii\ has dropped by a factor of 0.28. This means that the ratio of \ave\Slongnii /\ave\Slongoiii\ has changed from 1.6 to 4.3, as shown in Table~\ref{tab:NIL}. This supports a conclusion that the ionization drops with increasing distance from \tC. Much of this change of ratio is due to the dramatic drop in \ave\Slongoiii\ SW of the \crossing\ while the \ave\Sshortoiii\ has increased five-fold from the \NEsuper\ values

\subsection{What do we learn from the Back-scattering?}
\label{sec:PredObs} 

\begin{deluxetable*}{lccc}
\floattable
\setlength{\tabcolsep}{0.02in}
\tabletypesize{\scriptsize}
\tablecaption{Predicted and Observed Back-scattering Velocities*
\label{tab:PredObs}}
\tablewidth{0pt}
\tablehead{
\colhead{Velocity Component}&
\colhead{~~Predicted (\Vscat -\Vcomp)} &
\colhead{~~Observed (\Vscat -\Vcomp) } &
\colhead{\Sscat /\Scomp } 
}
\startdata
{\bf NE-Sector} & & & \\
\Vlongnii                   &  16$\pm$2        & 16$\pm$3                      & 0.11$\pm$0.04   \\
\Vlongoiii                   & 22$\pm$4             &21$\pm$4             & 0.07$\pm$0.03    \\    
\Vshortoiii                    & 44$\pm$4      &---**      & ---** \\  
\crossing & & & \\
\Vlongnii                   &  10$\pm$2        & 16$\pm$5                      & 0.08$\pm$0.05   \\
\Vlongoiii                   & 17$\pm$4             &21$\pm$2(25),18$\pm$3(16)     & 0.20$\pm$0.09(29),1.0$\pm$0.3(15)    \\    
\Vshortoiii                     &  43$\pm$3      & 34$\pm$3(29),29$\pm$3(15)     & 0.30$\pm$0.17(27),0.19$\pm$0.09(12) \\  
{\bf SW-Sector} & & & \\
\Vlongnii                   &  16$\pm$2        & 16$\pm$1                      & 0.13$\pm$0.03   \\
\Vlongoiii                   & 17$\pm$5             &19$\pm$2,15$\pm$2(9)     & 0.20$\pm$0.09(11),1.2$\pm$0.6(9)  \\    
\Vshortoiii                     &  44$\pm$4       & 31$\pm$2(11), 27$\pm$4(9)     &  0.33$\pm$0.11(11),0.16$\pm$0.10(9)\\  
\enddata
~*~All velocities are Heliocentric velocities in \kms.  

**No \Vshortoiii\ component is seen 

**Samples group around these values. Parentheses enclose the number of samples within each subgroup. 
\end{deluxetable*}

We can apply the methods described in Sections~\ref{sec:deltaV} and \ref{sec:Albedo} to the interpretation of the regions immediately to the NE of the \crossing\ (the NE-Sector) , the \crossing\ itself, and the regions immediately to the SW (the SW-Sector). For this we have employed the spectra in the profiles, as described in Section~\ref{sec:AnalProfiles}. The results of our observations and predictions are shown in Table~\ref{tab:PredObs}.

As noted in Section~\ref{sec:Albedo}, the ratio \Sscat /\Scomp\ should be much less than unity and this can be used to verify that one has correctly identified the source of the \Vscat\ component. This condition is satisfied for all of the \nii\ samples and for most of the \oiii\ samples. However, in the \crossing\ we see that the more numerous \Vlongoiii\ component  satisfies the condition while in \Vshortoiii\ the less numerous do. The same pattern applies in the SW-Sector (although the population of the groups are about equal).  

It is informative to compare the observed velocity difference (\Vscat -\Vcomp) with the predicted value. For a flat-on region the observed \Vcomp\ will be 
\Vpdr -{\bf V$_{expansion}$} from which the photo-ionization flow velocity for each emission-line can be calculated. We have used \Vpdr = 27 \kms\ and from this calculated a predicted 
\Vscat -\Vcomp . This value should decrease in a more highly tilted region. 

In \nii, the long component is dominant (c.f. Table~\ref{tab:AllniiSources}. The agreement of \Vscatnii -\Vlongnii\  observed and predicted values is good for the NE- and SW-Sectors, which are not highly tilted, and the difference in the \crossing\ indicates the complexity of the region. The \Sscatnii /\Slongnii\ values are always low enough to confirm that \Vlongnii\ is the source of the backscattered light.  

In the NE-Sector \Sshortoiii /\Slongoiii\ is small( to the point that \Vshortoiii\ is not detected). The observed and predicted \Vscatoiii -\Vlongoiii\ are in good agreement. 

In the \crossing , we see that the observed and predicted velocity difference is very different for \nii .  This probably reflects the complexity of this region owing to the many smaller regions with large tilts.

Also in the \crossing\ we see that a larger sample satisfies the \Sscatoiii -\Slongoiii\ requirement, but that its observed velocity different agree's less well than the smaller group. However, the probable errors are such that it may be indistinguishable. Neither \Vshortoiii\ group agrees with the predictions. Again, this
indicates the complexity of the \crossing.

In the SW-Sector the observed and predicted velocity differences for \nii\ are excellent. Again the \Vlongoiii\ component breaks down into favored and unlikely scatters, but both components agree equally well with the predictions. Like the \crossing , the observed and predicted values are highly discrepant.

In summary, we see that in \nii\ the observed and predicted velocity differences are good in the NE and SW-Sectors, but poor in the complex \crossing.
In \oiii\ we see that the agreement is good in the NE and SW-Sectors, but in the \crossing\ there are no good agreements.

\subsection{A different \Vevap\  velocity for \oiii }
\label{sec:Vevap}

The predicted values of \Vscat -\Vcomp\ in the previous section were calculated from inferred values of \Vevap\ and that the velocity separation would be twice \Vevap. The argument can be reversed to use the observed velocity difference to derive \Vevap . This tells us that \Vevapnii\ is the same in all three Supergroups, about 8$\pm$1 \kms.  Using the lower \Sscatoiii /\Slongoiii\ groups gives 
\Vevapoiii = 10$\pm$1 \kms. The slightly larger value for \oiii\ is consistent with the idea \citep{hen05}that the \oiii\ emitting zone is further from the MIF and thus having been subjected to more acceleration.

    \section{Conclusions}
    \label{sec:conclusions}
    
$\bullet$ The \nii\ emission arises from a narrow layer along an ionization front. The \Vlongnii\ component is always much brighter than the \Vshortnii\ component.
In the region nearest \tC\ it clearly lies along the MIF and in the \crossing\ along the ionized surface of the \cloud\ that faces the observer. To the SW its location is uncertain as it could be either over the main body of the \cloud\ or on the MIF of the nebula beyond the embedded \cloud. Strong backscattering of its radiation indicates that it is always close to a dusty PDR.

$\bullet$ The \Vshortnii\ component arises from the NIL, the layer of gas lying on the observer's side of \tC\ and the \cloud. Its surface brightness is the same in the \NEsuper\ and the \SWsuper, even though the latter is much more distant in the plane of the sky. The surface brightness is unexpectedly much higher in the \CRsuper. Its velocity is almost the same when sweeping from NE to SW, but there is a possible increase 
over the \CRsuper. If real, this would indicate that the NIL is closer to \tC\ than we calculated in Paper-I.

$\bullet$ The \Vlongoiii\ component arises from a thicker, more highly ionized region than \Vlongnii. Its velocity is almost constant when passing through the {\bf Supergroups}. Like \Vlongnii , the \NEsuper\ emission arises from the MIF beyond \tC\ and the \CRsuper\ on the observer's side of the \cloud.  Again, the physical location is uncertain in the \SWsuper . Its surface brightness decreases monotonically with increasing distance from \tC. The lack of an increase in the \CRsuper\ indicates that its emitting layer is thicker than the physical high point indicated by the \nii\ profiles. The \oiii\ profiles indicate that at the \crossing\ the surface begins to drop and beyond the \crossing\ \Vshortoiii\ is dominant. Within the \crossing\ we see wide variations in \Sshortoiii /\Slongoiii\ and note that the surface brightness variations suggest that the EUV radiation is being split between the \Sshortoiii\ and \Slongoiii\ emitting regions.  

$\bullet$ \Vshortoiii\ is essentially constant across the NE-SW sweep. Taken alone this would indicate that it always arises from the NIL.
However, the surface brightness of \Vshortoiii\ leaps in the \CRsuper\  and remains high into the \SWsuper. This behavior remains unexplained. 

$\bullet$ Within the peak of the rise upon which the \crossing\ is centered there is a highly tilted low ionization region facing the north.

$\bullet$ The \cloud\ produces changes to its SW, which is the direction of the hot gas giving rise to X-ray emission. However, there is no obvious link to the Outer Shell that covers the near side of the Extended Orion Nebula. 

$\bullet$ There is a pressing need to refine the NIL modeling approach that we presented in Paper-I and to apply it to various positions along the NE-SW sweep.

 \section*{acknowledgements}
 
The observational data were obtained from observations with the NASA/ESA Hubble Space Telescope,
obtained at the Space Telescope Science Institute (GO 12543), which is operated by
the Association of Universities for Research in Astronomy, Inc., under
NASA Contract No. NAS 5-26555; the Kitt Peak National Observatory and the Cerro Tololo Interamerican Observatory operated by the Association of Universities for Research in Astronomy, Inc., under cooperative agreement with the National Science Foundation; and the San Pedro M\'artir Observatory operated by the Universidad Nacional Aut\'onoma de M\'exico. 
We have made extensive use of the SIMBAD data base, operated at CDS, Strasbourg, France and its mirror site at Harvard University and NASA's Astrophysics Data System Bibliographic Services. 
GJF acknowledges support by NSF (1816537, 1910687), NASA (ATP 17-ATP17-0141), and STScI (HST-AR- 15018). 

\appendix
\section{A profile across the Ledge feature at the center of the Crossing}
\label{sec:DkArcProfile}

\begin{figure*}
 \includegraphics
[width=7.5in]	
 {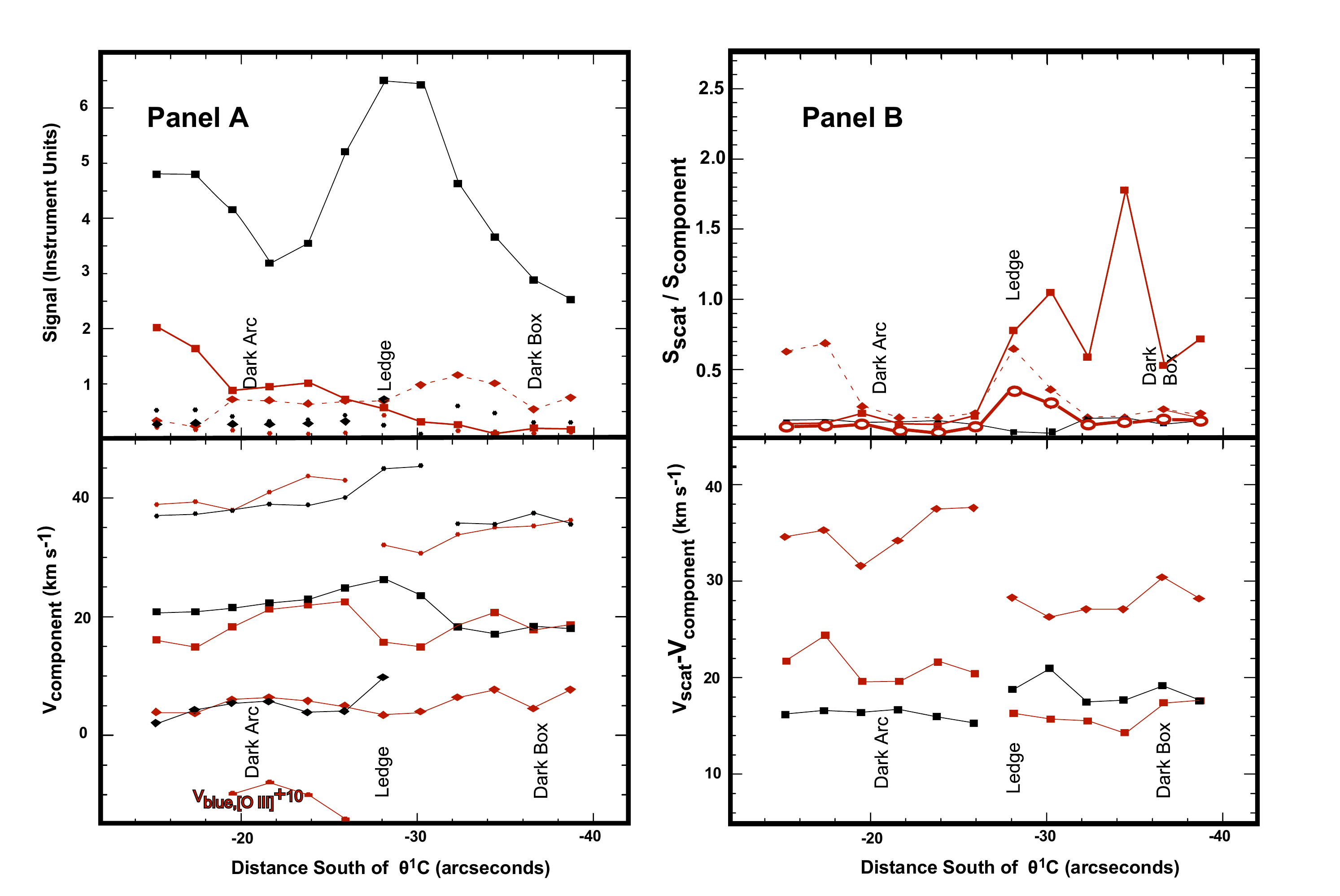}
\caption{
This figure is similar to Figure~\ref{fig:Profiles} except that it shows data from a profile at an RA displacement 38\arcsec\ west of \tC\ passing through the center of the \ledge\ feature in the center of the Crossing. The symbols mean the same as in Figure~\ref{fig:Profiles} but the red ellipses indicate \Sscatoiii /(\Sshortoiii +\Slongoiii). In Panel A Velocities and Signals are shown. Weak but clear {\bf V$\rm _{blue,[O~III]}$} components are also shown in their correct displacements but at velocities 10 \kms\ greater than measured. In Panel B scattered light ratios and velocity differences are shown.}
\label{fig:RA38profiles}
\end{figure*}

At the center of the 38\arcsec\ west profile lies a narrow  E-W feature within the larger feature previously called the West-jet. It does not have any measured radial and tangential velocities (\citep{ode15}, Paper-IV), thus arguing that the name originally assigned to it (the West-jet) is misleading.The small \Vlongnii\ peak there and the large increase in \Slongnii\ mean that it is a small ionized region seen nearly edge on, therefore a designation as the \ledge\ is more descriptive. The higher side of the \ledge\ lies to the south, otherwise there would be a shadowed zone in that direction. 

Figure~\ref{fig:RA38profiles} shows the results along the RA38 profile. Proceeding south in the upper portion of Panel A we see that there are dips in \Slongnii\ and \Slongoiii\ along the Dark Arc, with the minimum of \Slongoiii\ being slightly further north, at which point \Sshortoiii\ begins to increase.
The large rise in \Slongnii\ but continuation of a decreasing trend in \Slongoiii\ south of the Dark Arc is consistent with the much higher spatial resolution images in Figures~\ref{fig:DkArc658} and \ref{fig:DkArc502}. This indicates that the \oiii\ emitting layer is thicker that the size of the Dark Arc in the plane of the sky.  \Slongoiii\ decreases south of the \ledge\ while \Sshortoiii\ remains strong, indicating that the \Sshortoiii\ component has become dominant. 

In the lower portion of Panel A one sees that the \Vlongnii\ component slowly increases until the \ledge\ is reached. This can be interpreted as an increasing tilt of the ionized layer removing more of the photoevaporation flow velocity and after the peak velocity at the \ledge, the drop in \Vlongnii\ and \Vlongoiii\ indicate the south side of the \ledge\ is flatter than the north.  The disappearance of the \Vshortnii\ component south of the \ledge\ indicates that the gas that had been producing it is no longer there, probably by having become more ionized.  

In the lower portion of Panel B we see that \Vscatnii -\Vlongnii\ is about the same ( 16$\pm$1 \kms) as found in Table~\ref{tab:AllData} for the \crossing\ and the slight shift to a larger value south of the \ledge\ is consistent with this region being flatter.  
In \oiii\ the velocity is again similar to the \crossing\ value in Table~\ref{tab:AllData} (20$\pm$3 \kms) but it is unclear why the difference decreases to the south, although this may be due to the \Vlongoiii\ feature becoming weak there. 
Similarly, the high values in \Sscatoiii /\Sshortoiii\ (upper portion of Panel B) occur in the north, where \Sshortoiii\ is weak and becomes appropriately low in the south where \Sshortoiii\ is strong. This means that even in \oiii\ the \Vscatoiii\ is backscattered  light from the nearest strong emitting layer. This is illustrated by the series of data-points giving the \Sscatoiii /(\Sshortoiii +\Slongoiii) values. The local rise in the \ledge\ samples probably means that our simple backscattering model breaks down there.

In the upper panel of the upper portion of Panel B we see that the ratio \Sscatnii /\Slongnii\ is always small, as one would expected from \Vlongnii\ radiation being backscattered from a nearby PDR.  Again, \oiii\ is more complex. However, in the north region, where \Vlongoiii\ is strongest, the \Sscatoiii /\Slongoiii\ values are small-again like backscattering. Then the ratio becomes impossibly large to the south, indicating that it is not the source of \Vscatoiii ; but, this is the region where \Vlongoiii\ has become weaker than \Vshortoiii. 

In the lower portion of Panel A we show the locations of four very blue (about -20 \kms) and weak (about 1$\%$ of \Slongoiii ) \oiii\ velocity components. 

The averaged results differ by no statistically significant amounts from the lower resolution profile results given under the \crossing\ heading in Table~\ref{tab:AllData}.
 There are two surprising results here; that \nii\ strong \ledge\ shows no difference in its radial velocity from that of the surrounding nebula and that the dark features do not contain velocity differences.


\end{document}